\documentclass[12pt,twoside,twocolumn]{article}

\usepackage{amsfonts,amsmath,amssymb,bm,abstract,balance,cite}

\title{\textbf{Reply to Comment on `Formation of bound states of electrons in
spherically symmetric oscillations of plasma'}}

\author{Maxim Dvornikov
\\
\small{Departamento de F\'{i}sica, Universidad T\'{e}cnica Federico Santa Mar\'{i}a,} \\
\small{Casilla 110-V, Valpara\'{i}so, Chile and} \\
\small{IZMIRAN, 142190, Troitsk, Moscow region, Russia;} \\
\small{E-mail: maxim.dvornikov@usm.cl}}

\date{}

\begin{document}

\twocolumn[\maketitle
\begin{onecolabstract}
I reply here to the comment of Dr Shmatov on my recent work and
demonstrate the invalidity of his criticism of the classical
physics description of the formation of bound states of electrons
participating in spherically symmetric oscillations of plasma.
\\
PACS numbers: 52.35.Fp, 92.60.Pw, 74.20.Mn
\\
\end{onecolabstract}]

Nowadays there is a lack of a universally recognized theoretical
model of stable natural plasma structures existing in the
atmosphere~\cite{Ste99}. An approach to the description of such a
plasmoid based on radial oscillations of electron gas was recently
put forward in Refs.~\cite{Dvo01,Shm03,Dvo10}. Oscillations of
electrons were treated in both quantum and classical
frameworks~\cite{Dvo01} within the proposed model. Various
important phenomena, such as emission of high energy radiation,
which arise in spherically symmetric plasma oscillations, were
also predicted~\cite{Shm03}. Note that other theoretical
descriptions of natural plasmoids, including very exotic ones,
were reviewed in Ref.~\cite{BycGolDij10}.

In Ref.~\cite{Dvo10} I analyzed the possibility of the formation
of bound states of electrons participating in spherically
symmetric oscillations due to the exchange of ion acoustic waves.
Note that in Ref.~\cite{Dvo10} the dynamics of electron
oscillations was treated in frames of the classical
electrodynamics. In the comment on my work made by
Dr.~Shmatov~\cite{Shm10}, it was claimed that the classical
physics description adopted in Ref.~\cite{Dvo10} is inconsistent
with the numerical estimates presented in my paper, since the
typical energy of an electron participating in oscillations is
below the minimal kinetic energy $E_\mathrm{q}$ (see
Ref.~\cite{Shm10}) resulting from the Heseinberg uncertainty
principle.

I disagree with the statement of Ref.~\cite{Shm10} that classical
electrodynamics is invalid for the description of the bound state
formation. Indeed, to form a bound state the energy of the
effective attraction should be greater than the kinetic energy of
electrons (see, e.g., Ref.~\cite{NamAka85}). As in
Ref.~\cite{Dvo10} we can discuss singly ionized nitrogen plasma
with the background electron density $n_0 =
10^{15}\thinspace\text{cm}^{-3}$ and electron temperature $T =
10^5\thinspace\text{K}$, which corresponds to the typical plasma
of a gas discharge~\cite{Bog02}. Note that this value of $T$ is
different from that used in Ref.~\cite{Dvo10}. Taking the
amplitude of electron oscillations $a \sim 10^2
k_\mathrm{e}^{-1}$, where $k_\mathrm{e}$ is the Debye wave number,
and the distance between oscillating electrons $d \sim 10 a$ as
well as using Eq.~(16) from Ref.~\cite{Dvo10}, we get that the
effective attraction takes place if $|\omega_\mathrm{i} - \Omega|
\leq 10^{-4} \omega_\mathrm{i}$, where $\omega_\mathrm{i}$ is the
ion Langmuir frequency and $\Omega$ is the frequency of the
electron motion. Note that in this case the kinetic energy of
electrons prevails both the energy of their thermal motion
$\sim\text{several eV}$ and $E_\mathrm{q} \sim 10^{-18}\text{eV}$.

Thus I have demonstrated that the classical electrodynamics
description of the bound state formation of electrons
participating in radial oscillations is still valid although one
should choose the values of the parameters of the system, like
$T$, $d$, and $a$, different from those in Ref.~\cite{Dvo10}.
Nevertheless, I thank Dr.~Shmatov for pointing out in his
comment~\cite{Shm10} the unsuccessful choice of the parameters in
my work~\cite{Dvo10}.

Finally I mention that electron harmonic motion on the frequency
$\Omega < \omega_\mathrm{i}$ should not be necessarily treated as
forced oscillations as in Ref.~\cite{Dvo10}. It was demonstrated
in Ref.~\cite{Kuz76} that a plasma oscillation has to be
considered as a wave packet where both rapid and slow motions are
present since a monochromatic Langmuir wave is likely to be
unstable~\cite{Gol84}. Thus the results of Ref.~\cite{Dvo10} can
be applied to those electrons which participate in slow
oscillations.

\section*{Acknowledgments}
  This work has been supported by CONICYT (Chile) through Programa
  Bicentenario PSD-91-2006. The author is thankful to
  G.~V.~Dvornikova for technical help with the preparation of the manuscript.

\balance

\end{document}